\documentclass{foldiclp}
\usepackage{mathptmx}
\usepackage{algorithm,algpseudocode}
\newtheorem{exmp}{Example}[section]
\newtheorem{theorem}{Theorem}
\algnewcommand\algorithmicinput{\textbf{Input:}}
\algnewcommand\Input{\item[\algorithmicinput]}
\algnewcommand\algorithmicoutput{\textbf{Output:}}
\algnewcommand\Output{\item[\algorithmicoutput]}
\algnewcommand\algorithmicforeach{\textbf{for each}}
\algdef{S}[FOR]{ForEach}[1]{\algorithmicforeach\ #1\ \algorithmicdo}

\newcommand\bcmdtab{\noindent\bgroup\tabcolsep=0pt%
  \begin{tabular}{@{}p{10pc}@{}p{20pc}@{}}}
\newcommand\ecmdtab{\end{tabular}\egroup}

  \title[A New Algorithm to Automate Inductive Learning of Default Theories]
        {A New Algorithm to Automate Inductive Learning of Default Theories }
        
  \author[F. Shakerin and E. Salazar and G. Gupta]
         {FARHAD SHAKERIN, ELMER SALAZAR, GOPAL GUPTA\\
         The University of Texas at Dallas, Texas, USA\\
         \email{fxs130430,ees101020,gupta@utdallas.edu}}

\jdate{March 2003}
\pubyear{2003}
\pagerange{\pageref{firstpage}--\pageref{lastpage}}
\doi{S1471068401001193}

\begin{document}
\nocite{*}

\label{firstpage}

\maketitle

  \begin{abstract}
    In inductive learning of a broad concept, an algorithm should be able to distinguish concept examples
from exceptions and noisy data. An approach through recursively finding patterns in exceptions turns
out to correspond to the problem of learning default theories. Default logic is what humans employ in
common-sense reasoning. Therefore, learned default theories are better understood by humans. In this paper, we present new algorithms to learn default theories in the form of non-monotonic logic programs. Experiments reported in this paper show that our algorithms are a significant improvement over traditional approaches based on inductive logic programming. Under consideration for acceptance in TPLP.

  \end{abstract}



\section{Introduction}
Predictive models produced by classical machine learning methods
are not comprehensible for humans because they are algebraic solutions 
to optimization problems such as risk minimization or data likelihood 
maximization. These methods do not produce any intuitive description
of the learned model. This makes it hard for users to understand and verify 
the underlying rules that govern the model. 
As a result, these methods do not produce any justification 
when they are applied to a new data sample. Also, extending the 
prior knowledge\footnote{In the rest of the paper we will use the term background 
knowledge to refer to prior knowledge \cite{Muggleton91}.}
in these methods requires the entire model 
to be re-learned. Additionally, no distinction is made between 
\textit{exceptions} and noisy data. 
\textit{Inductive Logic Programming} \cite{Muggleton91}, however,
is one technique where the learned model is in the form of logic programming rules 
(Horn clauses) that
are more comprehensible and that allows the background knowledge to be incrementally
extended without requiring the entire model to be relearned.
This comprehensibility of symbolic rules makes 
it easier for users to understand and verify the resulting model 
and even edit the learned knowledge. 

Given the background knowledge and a set of positive and negative examples,
ILP learns theories in the form of Horn logic programs.
However, due to the lack of negation-as-failure, Horn clauses are not sufficiently expressive for representation 
and reasoning when the background knowledge is incomplete. 

Additionally, ILP is not able to handle exception to general rules: it 
learns rules under the assumption that there are no exceptions to them.
This results in exceptions and noise being treated in the same manner. Often,
the exceptions to the rules themselves follow a pattern, and these exceptions 
can be 
learned as well. The resulting theory that is learned is a default theory, and in most
cases this theory describes the underlying model more accurately. It should be
noted that default theories closely model common sense reasoning as well \cite{Baral}. Thus, 
a default theory, if it can be learned, will be more intuitive and comprehensible
for humans. Default reasoning also allows us to reason in absence of information. 
A system that can learn default theories can therefore learn rules that can draw 
conclusions based on lack of evidence, just like humans.
Other reasons that underscore the importance of inductive learning of default theories can be
found in Sakama \cite{Sakama05} who also surveys other attempts in
this direction. 

As an example, suppose we want to learn the concept of flying ability of birds. 
We would like to learn the default rule that birds normally fly, as well as 
rules that capture exceptions, namely, penguins and ostriches are birds 
that do not fly. Current ILP systems will be thrown off by the exceptions and will not discover any general rule: they will just either enumerate all the birds that
fly or cover the positive examples without caring much about the falsely  covered negative examples. Other algorithms, such as FOIL, will induce rules that are non-constructive and thus not helpful or intuitive.

In this paper, we present two algorithms for learning default 
theories (i.e., non-monotonic logic programs), called FOLD (First Order Learner of Default) and FOLD-R, to handle categorical and numeric features respectively. Unlike traditional
ILP systems that learn standard logic programs (i.e., no negation is allowed),
our algorithms learn \textit{non-monotonic stratified logic programs} (that allow \textit{negation-as-failure}).
Our algorithms are an extension of the FOIL
algorithm \cite{Quinlan90} and
support both categorical 
and numeric features. Also, the FOLD and FOLD-R learning algorithms can learn recursive rules. Whenever needed, our algorithms introduce new 
predicates. The language bias \cite{Mitchell80} also contains arithmetic constraints 
of the form $\{ A \leq h, A \geq h\}$. The algorithms have been implemented and
tried on variety of datasets from the UCI repository. They have shown 
excellent results that are presented here as well.

The default theories that we learn using our algorithm, as well as the background
knowledge used, is assumed to follow the stable model semantics\footnote{\label{bknote}We assume that the background knowledge has exactly one stable model.}. Stable model
semantics, and its realization in answer set programming(ASP), provides an elegant
mechanism for handling negation in logic programming 
\cite{GelfondL88}.  We assume that the reader is familiar with
ASP and stable model semantics \cite{Baral}.

This paper makes the following contributions: We propose a novel concrete algorithm to learn default theories automatically in the absence of complete information. The proposed algorithm, unlike the existing ones, is able to handle the numeric features without discretizing them first, and is also capable of handling non-monotonic background knowledge. We provide both qualitative and quantitative results from standard UCI datasets to support the claim that our algorithm discovers more accurate as well as more intuitive rules compared to the conventional ILP systems. 

Rest of the paper is organized as follows: Section 2 formally defines the problem we tackle in this paper. Section 3 presents some background materials. Section 4 presents the FOLD algorithm to solve the problem. In section 5 we extend FOLD to handle numeric features. Section 6 presents the experiments and results. Section 7 discusses related research. Section 8 discusses our future research direction and finally in Section 9 we conclude.

\section{The Inductive Learning Problem}
The problem that we tackle in this paper is an inductive non-monotonic logic programming problem which can be formalized as follows:\\
\textbf{Given}
\begin{itemize}
  \item a background theory $\mathcal{B}$, in the form of a normal logic 
program, i.e, clauses of the form $h \leftarrow l_1,...,l_m,not \ 
l_{m+1},...,not \ l_n.$ where $h$ and $l_1,...,l_n$ are positive literals and $not$ 
denotes \textit{negation-as-failure (NAF)} with stable model semantics;
  \item two disjoint sets of grounded goal predicates 
$\mathcal{E^+}$,$\mathcal{E^-}$, known as positive and negative examples respectively;
  \item a hypothesis language of predicates $\mathcal{L}$ including function 
and atom free predicates. It also contains a set of arithmetic constraints of 
the form $\{A \leq h, A \geq h\}$ where $A$ is a variable and $h$ is a real 
number;
  \item a $covers(\mathcal{H},\mathcal{E},\mathcal{B})$ function, which returns the 
subset of $\mathcal{E}$ which is extensionally implied by the current 
hypothesis $\mathcal{H}$ given the background knowledge $\mathcal{B}$; 
  \item a $score(\mathcal{E^+},\mathcal{E^-},\mathcal{H},\mathcal{B})$ function, which 
specifies the quality of the hypothesis $\mathcal{H}$ with respect to 
$\mathcal{E^+},\mathcal{E^-},\mathcal{B}$;
\end{itemize}
\textbf{Find}
\begin{itemize}
\item a theory $\mathcal{T}$ for which 
$covers(\mathcal{T},\mathcal{E^+},\mathcal{B}) = \mathcal{E^+}$ and 
$covers(\mathcal{T},\mathcal{E^-},\mathcal{B}) = \emptyset$.
\end{itemize}

\section{Background}
Our algorithm to learn default theories is an extension of the FOIL algorithm \cite{Quinlan90}. FOIL is a top-down ILP system which follows a \textit{sequential covering} approach to induce a hypothesis. The FOIL algorithm is summarized in Algorithm ~\ref{algo:foil}. This algorithm repeatedly searches for clauses that score best with respect to a subset of 
positive and negative examples, a current hypothesis and a heuristic called \textit{information gain} (IG). 

\begin{algorithm}

\caption{Summarizing the FOIL algorithm}
\label{algo:foil}
\begin{algorithmic}[1]
\Input $goal, \mathcal{B,E^+,E^-}$ 
\Output 
Initialize $\mathcal{H} \gets \emptyset $
\While{\textbf{not}($stopping \ criterion$)}
	\State $c \gets (goal$ :- $ \ true.)$
	\While{\textbf{not}($stopping \ criterion$)}
		\For{all $ \ c' \in \rho (c)$}
        	\State $compute \ score(\mathcal{E^+},\mathcal{E^-},\mathcal{H} \cup \{c'\},\mathcal{B})$
    	\EndFor
    	\State let $\hat{c}$ be the $c' \in \rho(c)$ with the best score  
    	\State $c \gets \hat{c}$
    \EndWhile	
    \State add $\hat{c}$ to $\mathcal{H}$
    \State $\mathcal{E^+} \gets \mathcal{E^+} \setminus covers(\hat{c},\mathcal{E^+},\mathcal{B})$
    
\EndWhile

\end{algorithmic}
\end{algorithm}

The inner loop searches for a clause with the highest information gain using a general-to-specific hill-climbing search. To specialize a given clause $c$, a refinement operator $\rho$ under $\theta$-subsumption \cite{plotkin70} is employed. The most general clause is $p(X_1,...,X_n) \gets true$, where the predicate $p/n$ is the predicate being learned and each $X_i$ is a variable. The refinement operator specializes the current clause $h \gets b_1,...b_n .$ This is realized by adding a new literal $l$ to the clause yielding $h \gets b_1,...b_n,l$. The heuristic based search uses information gain. In FOIL, information gain for a given clause is calculated as follows 
\cite{Mitchell97}:
\begin{equation}
IG(L,R) = t\left(log_2 \frac{p_1}{p_1 + n_1} - log_2 \frac{p_0}{p_0+ n_0} 
\right)
\end{equation}
where $L$ is the candidate literal to add to rule $R$, $p_0$ is the number of 
positive examples implied by the rule $R$, $n_0$~is the number of negative examples implied by the rule $R$, 
$p_1$ is the number of positive examples implied by the rule $R+L$, $n_1$ is the number of negative examples implied by the rule $R+L$, $t$ is the number of positive examples implied by $R$ also 
covered by $R+L$.
FOIL handles negated literals in a naive way by adding the literal $not \ L$ to 
the set of specialization candidate literals for any existing candidate $L$. 
This approach leads to learning predicates that do not capture the concept accurately as shown in the following example.

\begin{exmp}
$ \mathcal{B}, \mathcal{E^+}$ are background knowledge and positive examples 
respectively with CWA and the concept to be learned is fly.

\begin{tabular}{cll}
$\mathcal{B}: $    & $bird(X) \leftarrow penguin(X).$ & \\
                   & $bird(tweety).$                  & $bird(et).$ \\
                   & $cat(kitty).$                    & $penguin(polly).$\\
$\mathcal{E^+}:$   & $fly(tweety).$                   & $fly(et).$ \\

\end{tabular}
\end{exmp}
The FOIL algorithm would learn the following rule:
\[ fly(X) \leftarrow not \ cat(X), not \ penguin(X).\]
Although this rule covers all the positives (tweety and et are not penguins and cats) and no negatives (kitty and polly do not satisfy the clause body), it still does not yield an intuitive rule. In fact, the correct theory in this 
example is  as follows: ``{\it Only birds fly but, among them there are 
exceptional ones who do not fly}''. It translates to the following Prolog rule:
\[
fly(X) \leftarrow bird(X), not \  penguin(X).
\]
which FOIL fails to discover.

\section{FOLD Algorithm}
The idea of our FOLD algorithm is to learn a concept as a default theory and possibly multiple exceptions. In 
that sense, FOLD tries first to learn the default by specializing a general 
rule of the form $goal(V_1,...,V_n) \leftarrow true.$ with positive literals. 
As in FOIL, each specialization must rule out some already covered negative 
examples without decreasing the number of positive examples covered 
significantly. Unlike FOIL, no negative literal is used at this stage. Once 
the IG becomes zero, this process stops. At this point, if some negative 
examples are still covered, they must be either noisy data samples or 
exceptions to the so far learned rule. As 
\cite{Srinivasan96} discuss, there is no pattern 
distinguishable in noise, whereas, in exceptions, there may exist a pattern that can 
be described using the same language bias. This can be viewed as a  subproblem to (recursively) find the rules governing a set of negative examples. To achieve that aim, FOLD 
swaps the current positive and negative examples and recursively calls the FOLD 
algorithm to learn the exception rule(s). Each time a rule is discovered for 
exceptions, a new predicate $ab(V_1,...,V_n)$ is introduced. To avoid name 
collision, FOLD appends a unique number at the end of the string $ab$ to 
guarantee the uniqueness of the invented predicates.

In case of noisy data or in the presence of uncertainty due to the lack of 
information, it turns out that there is no pattern to learn. In such cases, 
FOLD enumerates the positive examples for two purposes: first, this is 
essential for the training algorithm to converge, second, it helps to detect 
noisy data samples.

Algorithm ~\ref{algo:fold} shows a high level implementation of the FOLD algorithm. In lines 
1-8, function FOLD, serves as the FOIL outer loop. In line 3, FOLD starts with 
the most general clause (e.g. $fly(X) \leftarrow true.$). In line 4, this clause 
is refined by calling the function $SPECIALIZE$. In lines 5-6, set of positive 
examples and set of discovered clauses are updated to reflect the newly 
discovered clause. In lines 9-29, the function $SPECIALIZE$ is shown. It serves 
as the FOIL inner loop. In line 12, by calling the function ADD\_BEST\_LITERAL 
the ``best'' positive literal is chosen and the best IG as well as the 
corresponding clause is returned. In lines 13-24, depending on the IG value, 
either the positive literal is accepted or the EXCEPTION function is called. 
If, at the very first iteration, IG becomes zero, then a clause that just 
enumerates the positive examples is produced. A flag called $just\_started$ 
is used to differentiate the first iteration. In lines 26-27, the sets of positive and negative 
examples are updated to reflect the changes of the current clause. In line 19, 
the EXCEPTION function is called while swapping the 
$\mathcal{E^+},\mathcal{E^-}$.    

In line 31, we find the ``best'' positive literal that covers more positive 
examples and fewer negative examples. Again, note the current positive examples 
are really the negative examples and in the EXCEPTION function, we try to find the 
rule(s) governing the exception. In line 33, FOLD is recursively called to 
extract this rule(s). In line 34, a new $ab$ predicate is introduced and in 
lines 35-36 it is associated with the body of the rule(s) found by the 
recurring FOLD function call in line 33. Finally, in line 38, default and 
exception are attached together to form a single clause.

The FOLD algorithm, once applied to Example 3.1 yields the following clauses:\\
\begin{tabular}{ll}
$fly(X) \leftarrow bird(X), not \ ab0(X).$ & \\
$ab0(X) \leftarrow penguin(X).$ & \\
\end{tabular}

\begin{algorithm}

\caption{FOLD Algorithm}
\label{algo:fold}
\begin{algorithmic}[1]
\Input $goal, \mathcal{B,E^+,E^-}$ 
\Output  \Statex $D = \{c_1,...,c_n\}$ \Comment{defaults' clauses}
		 \Statex $AB = \{ab_1,...,ab_m\}$ \Comment{exceptions/abnormal 
clauses}
\Function{FOLD}{$\mathcal{E^+},\mathcal{E^-}$} 
\While{($size(\mathcal{E^+}) > 0 $)}
\State $c \gets (goal$ :- $ \ true.)$
\State $\hat{c} \gets \Call{specialize}{{c},{\mathcal{E^+}},{{\mathcal{E^-}}}}$
\State $\mathcal{E^+} \gets \mathcal{E^+} \setminus 
covers(\hat{c},\mathcal{E^+},\mathcal{B})$
\State $D \gets D \cup \{ \hat{c} \}$
\EndWhile 
\EndFunction
\Function{SPECIALIZE}{${c},{\mathcal{E^+}},{\mathcal{E^-}}$}
\State $just\_started \gets true$
\While{($size(\mathcal{E^-}) > 0 $)}
\State  $(c_{def}, \hat{IG}) \gets 
\Call{add\_best\_literal}{{c},{\mathcal{E^+}},{{\mathcal{E^-}}}}$
\If{$\hat{IG} > 0$}
	\State $\hat{c} \gets c_{def} $
\Else
	 \If{$just\_started$}
			\State $\hat{c} \gets enumerate(c,\mathcal{E^+})$
	 \Else	
			\State $\hat{c} \gets 
\Call{exception}{{c},{\mathcal{E^-}},{{\mathcal{E^+}}}}$
			 \If {$\hat{c} = null$}
						\State $\hat{c} \gets 
enumerate(c,\mathcal{E^+})$
					\EndIf	
	 \EndIf
\EndIf
\State $just\_started \gets false$
\State $\mathcal{E^+} \gets \mathcal{E^+} \setminus 
covers(\hat{c},\mathcal{E^+},\mathcal{B})$
\State $\mathcal{E^-} \gets \mathcal{E^-} \setminus 
covers(\hat{c},\mathcal{E^-},\mathcal{B})$

\EndWhile
\EndFunction

\Function{EXCEPTION}{${c_{def}},{\mathcal{E^+}},{\mathcal{E^-}}$}
\State  $\hat{IG} \gets 
\Call{add\_best\_literal}{{c},{\mathcal{E^+}},{{\mathcal{E^-}}}}$
\If{$\hat{IG} > 0$}
	\State $ c\_set \gets \Call{FOLD}{\mathcal{E^+},\mathcal{E^-}} $
	\State $ c\_ab \gets generate\_next\_ab\_predicate()$
	\ForEach {$c \in c\_set $}
		\State $AB \gets AB \cup \{ c\_ab $:-$ \ bodyof(c) \}$
	\EndFor
	\State $\hat{c} \gets (headof(c_{def}) $:-$ \ bodyof(c), 
\textbf{not}(c\_ab))$
\Else
	\State $\hat{c} \gets null$
\EndIf

\EndFunction

\end{algorithmic}
\end{algorithm}

Now, we illustrate how FOLD discovers the above set of clauses given 
$\mathcal{E^+} = \{tweety,et\}$ and $\mathcal{E^-} = \{polly,kitty\}$ and the 
goal $fly(X)$. By calling FOLD, in line 2 ``while'', the clause $fly(X) 
\leftarrow true.$ is specialized. In the $SPECIALIZE$ function, in line 12, the 
literal $bird(X)$ is picked to add to the current clause, to get the clause 
$\hat{c} = fly(X) \leftarrow bird(X)$ which happened to have the greatest IG 
among $\{bird,penguin,cat\}$. Then, in line 26-27 the following updates are 
performed: $\mathcal{E^+}=\{\}$,\ $\mathcal{E^-}=\{polly\}$. A negative example 
$polly$, a penguin is still covered. In the next iteration, $SPECIALIZE$ fails 
to introduce a positive literal to rule it out since the best IG in this case 
is zero. Therefore, the EXCEPTION function is called by swapping the 
$\mathcal{E^+}$, $\mathcal{E^-}$. Now, FOLD is recursively called to learn a 
rule for $\mathcal{E^+} = \{polly\}$, $\mathcal{E^-}=\{\}$. The recursive call 
(line 33), returns $fly(X) \leftarrow penguin(X)$ as the exception. In line 34 
a new predicate $ab0$ is introduced and in line 35-37 the clause $ab0(X) 
\leftarrow penguin(X)$ is created and added to the set of invented 
abnormalities namely, AB. In line 38, the negated exception (i.e not \ 
$ab0(X)$) and the default rule's body (i.e $bird(X)$) are compiled together to 
form the clause $fly(X) \leftarrow bird(X),not \ ab0(X)$.     

Note, in two different cases $enumerate$ is called. First, at very first 
iteration of specialization if IG is zero for all the positive literals. 
Second, when the $Exception$ routine fails to find a rule governing the negative 
examples. Whichever is the case, corresponding samples are considered as noise. 
The following example shows a learned logic program in presence of noise. In particular, it shows how $enumerate$ function in FOLD works: It generates clauses in which the variables of the goal predicate can be unified with each member of a list of the examples for which no pattern exists.
\begin{exmp}
Similar to Example 3.1, plus we have an extra positive example fly(jet) without any 
further information:\\
\begin{tabular}{cll}
$\mathcal{B}: $    & $bird(X) \leftarrow penguin(X).$ & \\
                   & $bird(tweety).$                  & $bird(et).$ \\
                   & $cat(kitty).$                    & $penguin(polly).$\\
$\mathcal{E^+}:$   & $fly(tweety). \ \ \ \  $     $fly(jet).$   & $fly(et).$ \\

\end{tabular}
\end{exmp}
FOLD algorithm on the Example 4.1 yields the following clauses:\\
\begin{tabular}{ll}
$fly(X) \leftarrow bird(X), not \ ab0(X).$ & \\
$fly(X) \leftarrow member(X,[jet]).$ & \\
$ab0(X) \leftarrow penguin(X).$ & \\

\end{tabular}\\
FOLD recognizes $jet$ as a noisy data. $member/2$ is a built-in predicate in 
SWI-Prolog to test the membership of an atom in a list.

Sometimes, there are nested levels of exceptions. The following example shows 
how FOLD manages to learn the correct theory in presence of nested exceptions.

\begin{exmp}
Birds and planes normally fly, except penguins and damaged planes that can't. 
There are super penguins who can, exceptionally, fly.

\begin{tabular}{cl}
$\mathcal{B}: $    & $bird(X) \leftarrow penguin(X).$                           
           \\
                   & $penguin(X) \leftarrow superpenguin(X).$                   
            \\
                   & $bird(a).  \ \ \ bird(b). \ \ \ penguin(c).  \ \ \ 
penguin(d).$       \\
                   & $superpenguin(e). \ \ \ superpenguin(f).     \ \ \ 
cat(c1).$            \\
                   & $plane(g). \ \ \ plane(h). \ \ \ plane(k). \ \ \ 
plane(m).$			\\
                   & $damaged(k). \ \ \ damaged(m).$				
						\\
$\mathcal{E^+}:$   & $fly(a). \ \ \ fly(b). \ \ \ fly(e).$			
		                \\
				   & $fly(f). \ \ \ fly(g). \ \ \ fly(h).$	
								\\	

\end{tabular}
\end{exmp}
FOLD algorithm learns the following theory:\\
\begin{tabular}{ll}
$fly(X) \leftarrow plane(X), not \ ab0(X).$& \\
$fly(X) \leftarrow bird(X), not \ ab1(X).$ & \\
$fly(X) \leftarrow superpenguin(X).$& \\
$ab0(X) \leftarrow damaged(X).$ & \\
$ab1(X) \leftarrow penguin(X).$ & \\

\end{tabular}

\begin{theorem}
\label{thm:termination}
The FOLD algorithm terminates on any finite set of examples.

\begin{proof}
It suffices to show that the size of $\mathcal{E^+}$ on every iteration of the FOLD function decreases (at line 5) and since $\mathcal{E^+}$ is a finite set, it will eventually becomes empty and the while loop terminates. Equivalently, we can show that every time the SPECIALIZE function is called, it terminates and a clause $\hat{c}$ that covers a non-empty subset of $\mathcal{E^+}$ is returned. Inside the SPECIALIZE function, if $\mathcal{E^-}$ is empty, then the function returns its input clause and the theorem trivially holds. Otherwise, two cases might happen: First, SPECIALIZE produces a clause which enumerates $\mathcal{E^+}$ and covers no negative example. In such a case it returns immediately and the theorem trivially holds. Second, SPECIALIZE calls the EXCEPTION function which may lead to a chain of recursive calls on FOLD function. In this case it suffices to show that on a chain of recursive calls on FOLD, the size of function argument i.e. $\mathcal{E^+}$ decreases each time a new call to FOLD occurs. That's indeed the case because every time a literal is added to the current clause at line 12, it covers fewer negative examples from $\mathcal{E^-}$ which in turn becomes the new $\mathcal{E^+}$ as the EXCEPTION function and subsequently the FOLD function is recursively called. Therefore, on consecutive calls to FOLD function, the size of input argument $\mathcal{E^+}$ is decreased until it eventually terminates.  
\end{proof}

\end{theorem}

\begin{theorem}
The FOLD algorithm always learns a hypothesis that covers no negative example (soundness).

\begin{proof}
It follows from the Theorem \ref{thm:termination} that every loop in the algorithm (and, subsequently the algorithm) terminates. In particular, the while loop inside the function SPECIALIZE terminates as soon as the negated loop condition (i.e., the number of negative examples covered by the rule being discovered equals zero) starts to hold. Since, for every learned rule, no negative example is covered, it follows that the FOLD algorithm learns a hypothesis which covers no negative example.
\end{proof}

\end{theorem}

\begin{theorem}
The FOLD algorithm always learns a hypothesis that covers all positive examples (completeness).
\begin{proof}
The proof is similar to the soundness proof.
\end{proof}
\end{theorem}

\section{Numeric Extension of FOLD}
ILP systems have limited application to data sets containing a mix of 
categorical and numerical features. A common way to deal with numerical 
features is to discretize the data to qualitative values. This approach leads 
to accuracy loss and requires domain expertise. Instead, we adapt the approach 
taken in the well-known C4.5 algorithm \cite{Quinlan93}. This algorithm is ranked no. 1 in the 
survey paper ``{\it Top 10 algorithms in 
datamining}'', \cite{datamining08}. For a numeric 
feature A, constraints such as $\{ A \leq h, A > h \}$ have to be considered where the threshold $h$ is found by 
sorting the values of $A$ and choosing the split between successive values that 
maximizes the information gain. In our FOLD-R algorithm that we propose and describe next, we perform the same method for a set 
of operators $\{<,\leq \}$ and pick the operator and threshold which 
maximizes the information gain. Also, we need to extend the ILP language bias 
to support the arithmetic constraints. 

Unlike the categorical features for which we use {\it propositionalization}
\cite{Kramer01}, for numerical features, we define a predicate that 
contains an extra variable which always pairs with a constraint. For example to 
extend the language bias for a numeric quantity ``age'' we could define predicates of the form $age(a,b)$ in the background knowledge, and the candidate to 
specialize a clause might be as follows: $age(X,N), N \leq 5$. However, the predicate $age/2$ never appears without the corresponding constraint.    

Algorithm ~\ref{algo:foldr} illustrates the high level changes made to FOLD, in order to 
obtain the FOLD-R algorithm. The function $test\_categorical$, as before, chooses the best 
categorical literal to specialize the current clause. The function 
$test\_numeric $ chooses the best numeric literal as well as the best 
arithmetic constraint and threshold with the highest $IG$. In line 5, if 
neither one leads to a positive $IG$, EXCEPTION is tried. If exception also fails, then $enumerate$ is called. Otherwise, $IG$s are compared and whichever is greater, the corresponding clause is chosen as the specialized clause of the current iteration.

\begin{algorithm}
\caption{FOLD-R Algorithm, Specialize function. Other functions are the same as 
in FOLD}
\label{algo:foldr}
\begin{algorithmic}[1]
\Function{SPECIALIZE}{${c},{\mathcal{E^+}},{\mathcal{E^-}}$}
\While{($size(\mathcal{E^-}) > 0 $)}
\State  $(\hat{c_1}, \hat{IG_1}) \gets 
test\_categorical(c,\mathcal{E^+},{\mathcal{E^-}})$
\State  $(\hat{c_2}, \hat{IG_2}) \gets 
test\_numeric(c,\mathcal{E^+},{\mathcal{E^-}})$
\If{$\hat{IG_1} = 0 \ \& \ \hat{IG_2} = 0$}
	\State $\hat{c} \gets \Call{exception}{{c},				
			{\mathcal{E^-}},{{\mathcal{E^+}}}}$
	 \If {$\hat{c} = null$}
						\State $\hat{c} \gets 
enumerate(c,\mathcal{E^+})$
	 \EndIf	
\Else
	\If {$\hat{IG_1} \geq \hat{IG_2}$}
		\State $ \hat{c} \gets \hat{c_1}$
	\Else
		\State $ \hat{c} \gets \hat{c_2}$	  
	\EndIf 
	
\EndIf
\State $\mathcal{E^+} \gets \mathcal{E^+} \setminus 
covers(\hat{c},\mathcal{E^+},\mathcal{B})$
\State $\mathcal{E^-} \gets \mathcal{E^-} \setminus 
covers(\hat{c},\mathcal{E^-},\mathcal{B})$

\EndWhile
\EndFunction

\end{algorithmic}
\end{algorithm}

\begin{table}
\caption{Play Tennis data-set, Numeric version }
\label{table:tennis}
\begin{tabular}{ccccc} 
 \hline
 Outlook & Temperature & Humidity & Wind & PlayTennis \\ [0.5ex] 
 \hline\hline
 sunny    & 75 & 70 & true  & Play \\ 
 sunny    & 80 & 90 & true  & Don't Play \\
 sunny    & 85 & 85 & false & Don't Play \\
 sunny    & 72 & 95 & false & Don't Play \\
 sunny    & 69 & 70 & false & Play \\ 
 overcast & 72 & 90 & true  & Play \\
 overcast & 83 & 78 & false & Play \\
 overcast & 83 & 65 & true  & Play \\
 overcast & 81 & 75 & false & Play \\
 rain     & 71 & 80 & true  & Don't Play \\
 rain     & 65 & 70 & true  & Don't Play \\ 
 rain     & 75 & 80 & false & Play \\
 rain     & 68 & 80 & false & Play \\
 rain     & 70 & 96 & false & Play \\[1ex] 
 \hline
\end{tabular}
\end{table}
\begin{exmp}
Table \ref{table:tennis} adapted from \cite{Quinlan93} is a dataset with numeric 
features ``temperature'' and ``humidity''. ``Outlook'' and ``Windy'' are 
categorical features. Our FOLD-R algorithm, for the goal $play(X)$,  and positive examples shown as records with label ``Play'', and negative examples shown as records with label ``Don't Play'' outputs the following clauses:

\begin{tabular}{ll}
$ play(X) \leftarrow overcast(X).$ & \\
$ play(X) \leftarrow temperature(X, A), A \leq 75,not \ ab0(X).$ & \\
$ ab0(X) \leftarrow windy(X),rainy(X).$ & \\
$ ab0(X) \leftarrow humidity(X, A), A \geq 95,sunny(X).$ & \\
\end{tabular}
\end{exmp}
FOLD-R results suggest an abnormal day to play is either a rainy and windy day 
or a sunny day with humidity above 95\%.

\section{Experiments and Results}
This section presents the results obtained with FOLD-R algorithm on some of the standard UCI datasets. To conduct the following experiments, we implemented the algorithm in Java. We used Prolog queries to process the background knowledge (the background knowledge is assumed to be represented as a standard Prolog program). For performing information gain computations and CWA generation 
of negative examples, we made use of the JPL library \cite{jpl} which interfaces SWI-Prolog version 7.1.23-34 \cite{swiweb} with Java. 
Our intention here is to investigate the quality of discovered rules both in terms of their 
accuracy and the degree to which they are consistent with the common sense understanding 
from the underlying concepts. To measure the accuracy, we implemented 10-fold cross-validation 
on each dataset and the mean of calculated accuracy is represented while the standard deviation for all the datasets were 5 percent or lower. At present, we are not greatly interested in the running 
time and/or space complexity of the algorithm: this will be subject of future research. All the learning tasks were performed using a 
PC with Intel(R) Core(TM) i7-4700HQ CPU @ 2.40GHz and 8.00 GB RAM. Execution times ranges from a few seconds to a few minutes. The bottleneck is the function that sorts the 
numeric values to pick the best threshold and operator. 
There are solutions to get around this such as \cite{Catlett91}. Table~\ref{table:accuracy} reports the execution time of FOLD-R on our benchmarks. The algorithm works much faster when no numeric feature is present in the dataset. It should be noted that calls to JPL add significant overhead to the algorithm's execution time.

\noindent {\textbf{Labor Relations:} The data includes a set of contracts which depending on their features 
(16 features) are classified as good or bad contracts. 
The following set of clauses for a good contract are discovered by FOLD-R:

\begin{tabular}{ll}
$good\_cont(X) \leftarrow wage\_inc\_f(X,A), A > 2, not \ ab0(X).$& \\
$good\_cont(X) \leftarrow holidays(X,A), A > 11.$ & \\
$good\_cont(X) \leftarrow hplan\_half(X),pension(X).$ & \\
$ab0(X) \leftarrow no\_long\_disability\_help(X).$ & \\
$ab0(X) \leftarrow no\_pension(X).$ & \\
\end{tabular}
According to the first rule, a contract with 2 percent wage increase (default) 
is a good contract except when the employer does not contribute in a possible 
long-term disability and a pension plan. According to the second rule, a 
contract with holidays above 11 days is also good. And, finally, if employer 
contributes in half of the health plan and entire pension plan, the contract is good.

\noindent
{\textbf{Mushroom:} This dataset includes descriptions of different species of mushrooms 
and their features which are used to classify whether they are poisonous or 
edible. The following set of clauses for a poisonous mushroom is discovered 
by FOLD:

\begin{tabular}{ll}
$poisonous(X) \leftarrow ring\_type\_none(X).$& \\
$poisonous(X) \leftarrow spore\_print\_color\_green(X).$& \\
$poisonous(X) \leftarrow gill\_size\_narrow(X),not \ ab2(X).$ & \\
$poisonous(X) \leftarrow odor\_foul(X).$& \\
$ab0(X) \leftarrow population\_clustered(X).$ & \\
$ab0(X) \leftarrow stalk\_surface\_below\_ring\_scaly(X).$ & \\
$ab1(X) \leftarrow stalk\_shape\_enlarging(X).$ & \\
$ab2(X) \leftarrow gill\_spacing\_crowded(X),not \ ab1(X).$ & \\
$ab2(X) \leftarrow odor\_none(X), not \ ab0(X).$ & \\
\end{tabular}
Note, the induced theory has nested exceptions. This nesting happens as a result of finding patterns for negative examples, which makes the FOLD algorithm perform more recursive steps until no covered negative example is left. 

\noindent Table \ref{table:accuracy} compares the accuracy of FOLD-R algorithm against that of ALEPH \cite{aleph}. The examples have been picked from well-known standard datasets for some of which ALEPH exhibits low test accuracy. In most cases, FOLD-R accuracy outperforms ALEPH. The experiments suggest that when absence of a particular feature value plays a crucial role in classification, our algorithm shows a meaningful higher accuracy. This follows from the fact that the classical ILP algorithms only make use of existent information as opposed to \textit{negation-as-failure} in which a decision is made based on the absence of information. As an example, in the credit-j dataset, our algorithm generates 4 rules with abnormality predicates. These rules cover positive examples which without abnormality predicates would have remained uncovered. However, in Bridges and Ecoli where no abnormality predicate is introduced by our algorithm, both ALEPH and FOLD-R end-up with almost the same accuracy. 

Even in cases where no improvement over accuracy is achieved, our default theory approach leads to simpler and more intuitive rules. As an example, in case of Mushroom, other ILP systems, including ALEPH and FOIL, would produce 9 rules with 2 literals each in the body to cover all the positives, while our FOLD algorithm, produces 3 single-literal rules and 1 rule with 2 literals in which the second literal takes care of the exceptions.

\begin{table}
\begin{tabular}{lcccc} 
\hline
dataset & size & ALEPH accuracy(\%) & FOLD-R accuracy(\%) & FOLD-R execution time(s) \\ [0.5ex] 
\hline\hline
Credit-au & 690 & 82 & 83 & 67\\
Credit-j & 125 & 53 & 81 & 20\\
Credit-g & 1000 & 70.9 & 78 & 87\\ 
Iris & 150 & 85.9 & 95 & 1.3\\ 
Ecoli & 336 & 91 & 90 & 6.1\\ 
Bridges & 108 & 89 & 90 & 0.8\\  
Labor & 57 & 89 & 94 & 0.4\\  
Acute(1) & 34 & 100 & 100 & 0.3\\
Acute(2) & 34 & 100 & 100 & 0.3\\
Mushroom & 7724 & 100 & 100 & 11.4 \\
\hline	
\end{tabular}
\caption{FOLD-R evaluation on UCI benchmarks }
\label{table:accuracy}
\end{table}

\section{Related Work}

Sakama in 
\cite{Sakama05} discusses the necessity of having a non-monotonic language bias to perform induction for default reasoning. It surveys some of the proposals directly adapted from ILP, like \textit{inverse resolution} \cite{MuggletonB88} and \textit{inverse entailment} \cite{Muggleton95} then he explains why these are not applicable to non-monotonic logic programs. Sakama then introduces an algorithm to induce rules from \textit{answer sets} which generalizes a rule from specific grounded rules in a bottom-up fashion. His approach in some cases yields premature generalizations that produces redundant negative literals in the body of the rule and therefore over-fitted to the training data. The following example illustrates what Sakama's algorithm would produce:
\begin{exmp}
\begin{tabular}{cll}
$\mathcal{B}: $    & $bird(X) \leftarrow penguin(X).$ & \\
                   & $bird(tweety).$                  & $bird(et).$ \\
                   & $bear(teddy).$ 	              & $crippled(et).$ \\
                   & $cat(kitty).$                    & $penguin(polly).$\\
$\mathcal{E^+}:$   & $fly(tweety).$                   &  \\

\end{tabular}
and the algorithm outputs the following rule:\\
\begin{tabular}{ll}
$fly(X) \leftarrow bird(X), not \ cat(X), not \ penguin(X), not \ bear(X), not \ crippled(X). $
\end{tabular}
in which some of the literals including not cat(X) and not bear(X) are redundant.
\end{exmp}
Additionally, since ASP systems have to 
ground the predicates to produce the answer set, introducing numeric data in background knowledge and also in the language bias is prohibited.

In a different line of research,~\cite{DimopoulosK95}, describes an algorithm to learn exception using the patterns in the negative examples. However, they don't make any use of NAF as the core notion of reasoning in the absence of complete information. Instead, their algorithm learns a hierarchical logic program, including classical negation, in which the order of rules prioritize their application and, therefore, it is not naturally compatible with standard Prolog. 

The idea of swapping positive and negative examples to learn patterns from negative examples has first been discussed in \cite{Srinivasan96} where a bottom-up ILP algorithm is proposed to specialize a clause after it has already been generalized and still covers negative examples. Similarly, \cite{InoueK97} proposes a bottom-up algorithm in two phases: First, producing monotonic rules via a standard ILP method, then specializing them by introducing negated literals to the body of the rule. In contrast, we believe our FOLD algorithm that takes a top-down approach learns programs that have a better fit, thanks to its support for numeric features and its better scalability. Lack of both are inherent problems in bottom-up ILP methods.

ALEPH \cite{aleph} is one of the most widely used ILP systems that uses a bottom-up generalization search to induce theories that covers the maximum possible positive examples. However, since the induced theory might be overly generalized, there is an option to refine the theory by introducing abnormality predicates that rule out negative examples by specializing an overly generalized rule. This specialization step is manual and unlike our algorithm, no automation is offered by ALEPH. Also, ALEPH does not support numeric features.

XHAIL \cite{Ray2009} is another non-monotonic ILP framework which integrates abductive, deductive and inductive forms of reasoning to form sets of ground clauses called kernel set and then generalize it to learn non-stratified logic programs.

In a different line of research,~\cite{Corapi2012} approaches the non-monotonic ILP problem by incorporating the power of modern ASP solvers to search for an optimal hypothesis among the generated so-called skeleton rules and a set of abducibles associated to them. The last two approaches do not scale up as the language bias grows.

\section{Future Work}

One advantage of our FOLD-R algorithm over the existing systems is the ability to handle non-monotonic background knowledge. Conventional ILP systems permit only standard Prolog programs to represent background knowledge. In contrast, our FOLD-R algorithm, once integrated with a top down \textit{answer set programming} system like s(ASP) \cite{MarpleG12}, will permit the background knowledge to be represented as an answer set program. The idea of leveraging the added expressiveness of non-monotonic logic-based background knowledge has been discussed in \cite{Seitzer97}. However, because Seitzer's method is based on grounding the background knowledge and then computing its answer sets, it is not scalable. Using a query-driven system allows our learning algorithm to be scalable. Extending our algorithms to allow non-monotonic logic-based background knowledge with multiple stable models is part of the future work.

Our eventual goal is to develop a unified framework for learning default theories: (i) in which we can learn hypotheses that are general answer set programs (i.e., these learned answer set programs may not be stratified), and (ii) that work with background knowledge that may be represented as a non-stratified answer set program. Also,  the optimality of the learned hypothesis due to greedy nature of information gain heuristic is not guaranteed and changing the search strategy from A* to better algorithms such as iterative deepening search is subject of future research.

\section{Conclusion}
In this paper, we introduced a new algorithm called FOLD to learn default theories. Next, we proposed FOLD-R which is an extension of FOLD to handle numeric features. Both the FOLD and the FOLD-R learning algorithms learn stratified answer set programs that allow recursion through positive literals. Our experiments based on using the standard UCI benchmarks suggest that the default theory, in most cases, describes the underlying model more accurately compared to conventional ILP systems. Default theories also happen to closely model common-sense reasoning. Thus, rules learned from FOLD and FOLD-R are more intuitive and comprehensible by humans. Unlike classical machine learning approaches which learn based on existing information, FOLD and FOLD-R are able to find patterns of information that is absent and express it via a default theory, i.e., as a non-monotonic logic program.  

\section*{Acknowledgement}

Authors are partially supported by NSF Grant No. 1423419.

\bibliographystyle{acmtrans}
\bibliography{foldiclpcite}

\label{lastpage}

\end{document}